\documentclass[aps,showpacs,prl,twocolumn,fleqn]{revtex4}
\usepackage{amssymb}
\usepackage{graphicx}
\usepackage{dcolumn}
\usepackage{bm}
\usepackage{amssymb,amsmath}
\usepackage{wasysym}
\usepackage{dsfont}
\usepackage{enumerate}
\usepackage{color}
\usepackage{bbm}
\usepackage{footnote}
\begin{document}

\title{Detection of high-power two-mode squeezing by sum-frequency generation}

\author{Faina Shikerman and Avi Pe'er}
\affiliation{Physics department and BINA center for nano-technology,
Bar Ilan University, Ramat Gan 52900, Israel}

\begin{abstract}

We introduce sum-frequency generation (SFG) as an effective physical
two-photon detector for high power two-mode squeezed coherent states
with arbitrary frequency separation, as produced by parametric
oscillators well above the threshold. Using a formalism of
``collective modes'', we describe both two-mode squeezing and degenerate squeezing on equal footing and derive simple relations between the input degree of squeezing and the measured SFG quadrature noise. We compare the proposed SFG detection to
standard homodyne measurement, and show advantages in robustness to detection inefficiency (loss of SFG photons) and acceptance bandwidth.
\end{abstract}

\pacs{42.50.Dv, 42.50.St, 42.65.Lm, 42.65.Ky}

\maketitle

Quantum mechanical squeezing - the reduction of fluctuations of an
observable below the standard quantum limit (SQL -
${1}/{\sqrt{N}}$, $N$ the total number of photons detected) at the
expense of increased fluctuations of the conjugate observable - is a
major resource in quantum information and quantum measurement. In
optics, squeezed states of light are key to methods of phase
measurement with precision beyond SQL, approaching the ultimate
Heisenberg limit $1/N$ \cite{Caves-_PRD_1981,Holland-Burnett_PRL_1993-Holland-Hall_PRA_1998}. Due to the
potential for a dramatic improvement in precision, sub-SQL
methods are appealing for metrology applications, such
as detection of gravitational waves
\cite{Goda-Mavalvala_Nature_2008}, precision spectroscopy
\cite{Polzik-Kimble_Appl_Phys_B_1992} and next generation atomic
clocks \cite{Oblak-Polzik_PRA_2005}.

Two major limitations exist for measurement of squeezing by
standard homodyne detection. The first is sensitivity to
photo-detection inefficiency, which reduces the usable
squeezing. Since squeezing is very sensitive to photon loss, and since detection inefficiency in standard homodyne is indistinguishable from loss, near unity detection efficiency is crucial to exploit the squeezing resource \cite{Takeno-Furusawa_OPEQ_2007,Vahlbruch-Schnabel_PRL_2008}. Another limitation of homodyne detection is detection bandwidth - while parametric down-conversion (PDC) can produce
two-mode squeezed states with arbitrary frequency separation,
the photo detectors bandwidth is restricted to several GHz at most.
Consequently, standard homodyne detection is effective only for
narrowband degenerate squeezing and cannot be used for two-mode or
broadband squeezing, especially above the oscillation threshold. Detection of
the phase correlation in two-mode squeezing requires a stable reference for the phase-sum, which is not easy to obtain for spectrally-separated mode pairs. Reports so far relied on delicate referencing to optical cavities and were limited
to few nanometer separation between the modes \cite{Villar-Nussenzveig_PRL_2005}.
\begin{figure}[h]
\includegraphics[width=8cm]{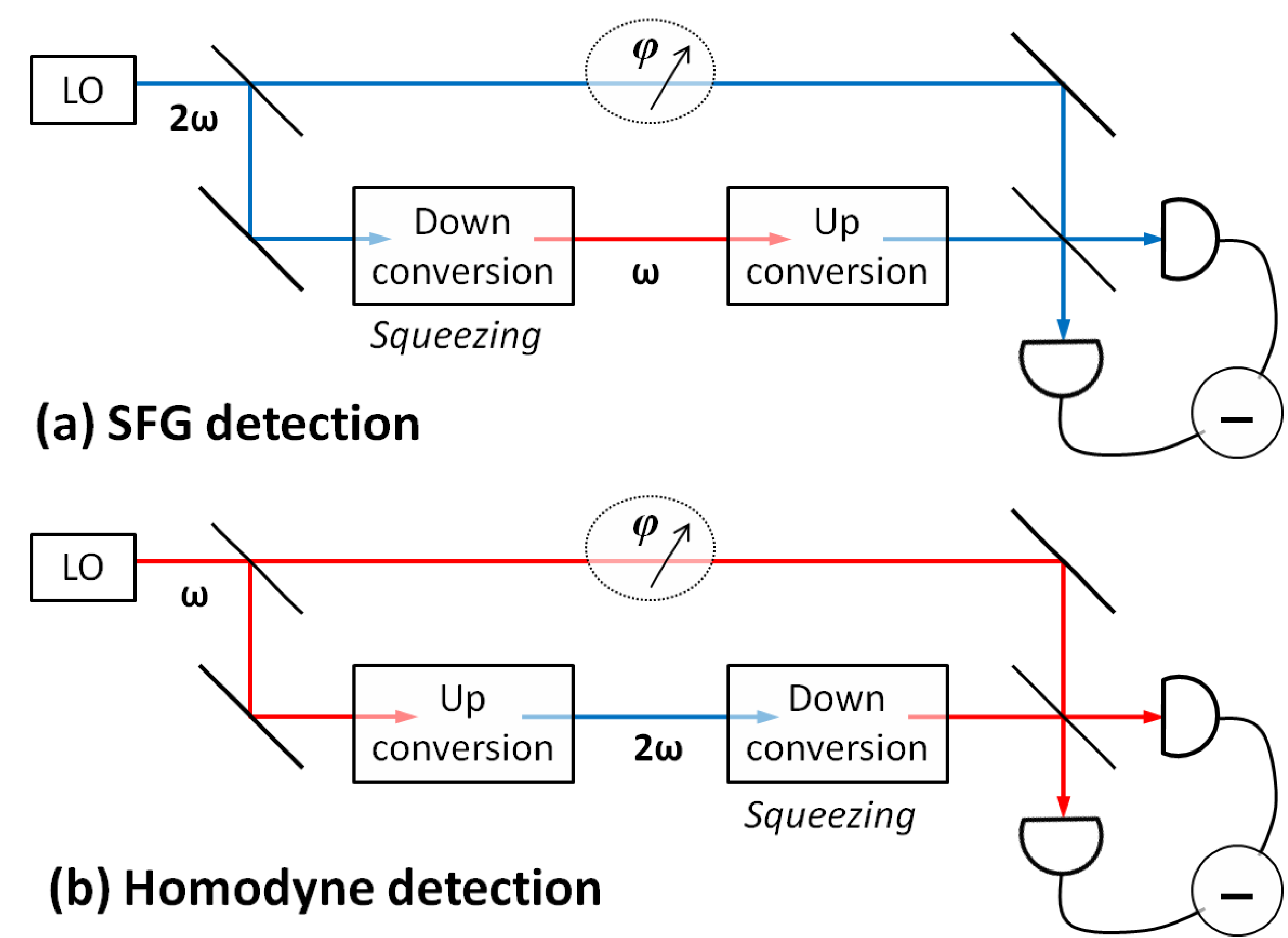}
\caption{\label{DetectionScheme} (Color online) (a) The proposed SFG
scheme for measurement of squeezing: a narrowband pump local oscillator (LO) at
frequency $2\omega$ is down-converted to generate squeezed light. To measure the obtained squeezing the light is first up-converted back to the pump frequency and the quadratures of the resulting SFG are measured by homodyning against the pump LO while varying its phase $\varphi$. This SFG scheme is
a symmetric inversion of the standard homodyne scheme shown in
(b), where a LO at $\omega$ is first frequency doubled to $2\omega$ and then down-converted to
generate squeezing, which is measured by homodyning against the LO.}
\end{figure}
\begin{figure}[h]
\includegraphics[width=8cm]{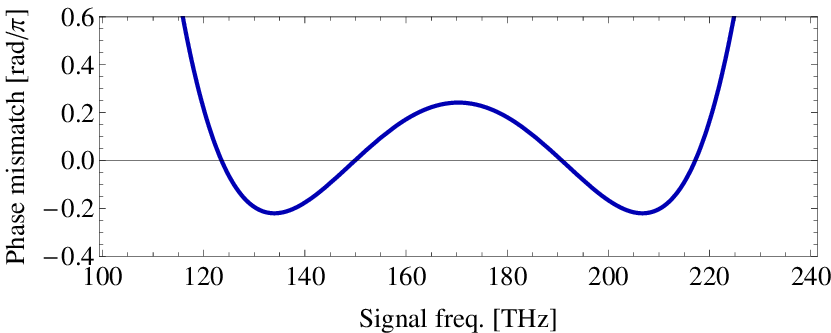}
\caption{\label{BroadbandPhaseMatching} Ultra-broad phase matching for SFG in a 1cm long PPKTP crystal with input near zero dispersion. $\!<\!0.25\pi$ phase mismatch is maintained over $\!\sim\!100THz$ input bandwidth for collinear SFG into 340THz (880nm).}
\end{figure}

We suggest a simple method to detect high-power two-mode squeezing, as produced by parametric oscillators above threshold \cite{SAT,Villar-Nussenzveig_PRL_2005,Peer-Friesem_PRA_2006}. The method, shown in Fig. \ref{DetectionScheme}, utilizes sum-frequency generation (SFG) as a detector of quantum correlation that is robust to detection inefficiency and accepts arbitrary frequency separation between the two modes. Previously, SFG was explored as an ultra-broadband two-photon detector in the low power regime of entangled bi-photons (squeezed vacuum), resolving simultaneously the tight time-difference and energy-sum correlation \cite{Dayan-Silberberg_PRL_2005,Peer-Silberberg_PRL_2005}. In the proposed scheme, the degree of squeezing of the input is deduced from analysis of the quadrature noise of the SFG output.

The Robustness of SFG detection to inefficiency is motivated by the fact that as opposed to homodyne, detection efficiency of the (double-frequency) SFG photons is easily distinguishable from loss of squeezed (fundamental frequency) photons. Furthermore, SFG involves annihilation of photon in pairs, which for low input depletion (low SFG efficiency) preserves the quantum correlation within the squeezed input. The ability to detect two-mode squeezing of arbitrary separation is motivated by the fact that the SFG phase is equal to the phase-sum of the input modes. Consequently, homodyne detection of the SFG phase against the original pump measures the inter-mode phase correlation regardless of their separation. The detection bandwidth with SFG is therefore only limited by phase matching, which can be very broad for type-I phase matching \cite{Dayan-Silberberg_PRL_2004,Dayan-Silberberg_PRL_2005,Peer-Friesem_CDMA_2004}, as shown in figure \ref{BroadbandPhaseMatching}.

In what follows, we consider a fully quantum model of the SFG setup of Fig. \ref{DetectionScheme}(a) and derive in linear response approximation analytic expressions relating the input degree of squeezing to the measured spectral variance of the SFG output. We employ the positive-P representation \cite{Prep,Gardiner} - a general method for treatment of quantum correlations -  to analyze SFG detection in realistic configurations, taking into account both loss and inefficiency. We start with a standard Hamiltonian \cite{LTS,OptActa28}
\begin{equation}
\hat{H}=\hat{H}_S +\hat{H}_R+\hat{V},\label{3H}
\end{equation}
where, setting $\hbar=1$,
\begin{equation}
\begin{split}
\hat{H}_S=&\sum_{\alpha=i,s,2}\omega_\alpha\hat{a}^\dag_\alpha
\hat{a}_\alpha
-i\frac{\chi}{2}\left(\hat{a}^\dag_2\hat{a}_i\hat{a}_s-\!\emph{h.c}\right)
\cr \quad &+i\left({\cal E}{\rm e}^{-i\omega_s
t}\hat{a}_s^{\dag}+{\cal E}^*{\rm e}^{-i\omega_i
t}\hat{a}_i^{\dag}-\!\emph{h.c}\right),
\end{split}\label{Hsfgsys}
\end{equation}
is the SFG intra-cavity Hamiltonian including the non-linear
interaction between the photon operators $\hat{a}_s,\hat{a}_i$ and
$\hat{a}_2$ of the signal, idler and SFG modes respectively, whose
frequencies obey energy conservation $\omega_i+\omega_s =\omega_2$.
The signal and the idler are driven by classical pumps ${\cal E}=|{\cal E}|{\rm e}^{-i\phi}$ and ${\cal E}^*$ that are complex conjugates to
reflect the classical correlation between them
\cite{Abram-Dolique_PRL_1986,Dayan-Silberberg_PRL_2004,Harris-_PRL_2007}.
$\hat{H}_R\!=\!\sum_k\!\omega_k\!\hat{b}^\dag_k\!\hat{b}_k$, is the
Hamiltonian of the loss reservoir of extra-cavity modes.
In rotating wave approximation, the cavity modes are coupled to the loss reservoir by \cite{Scully}
\begin{equation}
\hat{V}=\sum_{\alpha=i,s}\sum_k g_{\alpha k}\left(\hat{b}^\dag_k\hat{a}_\alpha+\!\emph{h.c}\right).\label{interactionTerm}
\end{equation}

To describe driving of a cavity by a quantum input, it is standard procedure to
separate the average classical field from the quantum fluctuations. Just as
driving by a coherent-state $|{\cal E}\rangle$ can be described as
coupling to a classical pump ${\cal E}$ accompanied by a vacuum
$|0\rangle$ input from the reservoir, we describe driving by a
\emph{squeezed} coherent state $|{\cal E},R{\rm
e}^{i\theta}\rangle$ ($R$ - the degree of squeezing) as a classical
pump ${\cal E}$ accompanied by a \emph{squeezed} vacuum
$|0,R{\rm e}^{i\theta}\rangle$ input from the reservoir. 

It is useful to transform the mode basis from the signal and idler
modes to collective modes, defined as
\begin{equation}
\hat{a}_\pm=\frac{1}{\sqrt{2}}\left(\hat{a}_s{\rm
e}^{i\phi}\pm\hat{a}_i{\rm e}^{-i\phi}\right).\label{apm}
\end{equation}
From classical analogy, the collective $\pm$ modes correspond to
a carrier at the center frequency $\omega_1\!=\!(\omega_i\!+\!\omega_s)/2\!=\!\omega_2/2$ modulated by a cosine or sine envelope at frequency $(\omega_s\!-\!\omega_i)/2$. Just like $a_s, a_i$, the collective
operators commute $\left[a_+,a_-\right]\!=\!\left[a_+^{\dag},a_-^{\dag}\right]\!=\!\left[a_+,a_-^{\dag}\right]\!=\!0$
and should not be confused with quadrature operators. Assuming $\langle\hat{b}_k\rangle=\langle\hat{b}^\dag_k\rangle=0$,
and using squeezed vacuum correlation functions \cite{Gardiner}
\begin{equation}
\begin{split}
&\langle\hat{b}^\dag_k\hat{b}_{k'}\rangle=N\delta\left(k,k'\right),\ \ \langle\hat{b}_k\hat{b}^\dag_{k'}\rangle=(N+1)\delta\left(k-k'\right),\\
&\langle\hat{b}^\dag_k\hat{b}^\dag_{k'}\rangle=\langle\hat{b}_k\hat{b}_{k'}\rangle^*=-M\delta\left(\omega_k,\omega_2-\omega_{k'}\right),
\end{split}
\end{equation}
we can substitute
$\hat{a}_{s,i}\!=\!\left(\hat{a}_+\pm\hat{a}_-\right){\rm e}^{\pm
i\phi}/{\sqrt{2}}$ (Eq.~(\ref{apm})) into the Hamiltonian
and apply general methods to construct a reduced master
equation in the Markovian limit \cite{Scully}:
\begin{equation}
\begin{split}
\frac{d\hat{\rho}}{dt}=&-i\left[\hat{H}_{int},\hat{\rho}\right]+\gamma_2\left(2\hat{a}_2\hat{\rho}\hat{a}_2^\dag-\hat{a}_2^\dag\hat{a}_2\hat{\rho}-\hat{\rho}\hat{a}_2^\dag\hat{a}_2\right)\\
&-{\cal N}\left[\hat{a}_+\right]+{\cal
M}\left[\hat{a}_+\right]-{\cal N}\left[\hat{a}_-\right]-{\cal
M}\left[\hat{a}_-\right],
\end{split}\label{GME}
\end{equation}
where
\begin{small}
\begin{equation}
\hat{H}_{int}\!=\!i\sqrt{2}|{\cal
E}|\left(\hat{a}^\dag_+\!-\!\hat{a}_+\right)\!-\!i\frac{\chi}{4}\left(
\hat{a}_2^\dag\hat{a}_+^2\!+\!\hat{a}_2^\dag\hat{a}_-^2\!-\!\emph{h.c}\right)
\label{Hint}
\end{equation}
\end{small}
is the system Hamiltonian $\hat{H}_S$ in the interaction picture representation, and the functions $\cal N, \cal M$ are defined as
\begin{small}
\begin{equation}
\begin{split}
{\cal N}\left[\hat{A}\right]\!\equiv\!\gamma_1(N+1)\left[\hat{A}^\dag\hat{A}\hat{\rho}-2\hat{A}\hat{\rho}\hat{A}^\dag+\hat{\rho}\hat{A}^\dag\hat{A}\right]
\cr \quad +\gamma_1N\left[\hat{A}\hat{A}^\dag\hat{\rho}-2\hat{A}^\dag\hat{\rho}\hat{A}+\hat{\rho}\hat{A}\hat{A}^\dag\right],\cr
{\cal M}\left[\hat{A}\right]\!\equiv\!
\gamma_1M\left[\hat{A}^2\hat{\rho}-2\hat{A}\hat{\rho}\hat{A}+\hat{\rho}\hat{A}^2\right]
+ \!\emph{h.c},
\end{split}\label{calNM}
\end{equation}
\end{small}
and $\gamma_1,\gamma_2$ are the decay rates of the signal $\pm$ idler and
the SFG modes respectively. The quantities $M$ and $N$,
appearing in Eqs.~(\ref{calNM}), characterize the input squeezing
and, in principle, can be varied independently
\cite{GardinerCollett}. For the ideal squeezed input, however,
$M\!=\!\sinh(2R){\rm e}^{-i\theta}/2$, $N\!=\!\sinh^2(R)$, and hence,
$|M|\!=\!\sqrt{N(N+1)}\!\approx\!N+1/2$. While arbitrary $\theta$
can be treated, we confine our analysis to real $M$
$(\theta\!=\!0,\pi)$ for brevity of expressions.

It is readily seen from Eq.~(\ref{Hint}) that with the definition
of collective modes (Eq.~(\ref{apm})) only the cosine envelope $\hat{a}_+$ is
driven directly, while the sine envelope $\hat{a}_-$ is neither
externally excited nor directly coupled to $\hat{a}_+$.
The only mechanism to populate $\hat{a}_-$ is by spontaneous down
conversion from the created SFG field, which is negligible if the SFG efficiency is low. The physical picture of the
two-mode SFG reduces therefore to that of a degenerate squeezing, apart from
the modulation of the $\omega_1$ frequency carrier by a cosine envelope. Discarding the terms
involving $\hat{a}_-,\hat{a}_-^\dag$ and setting
$\hat{a}_+\!=\!\hat{a}_1$, Eqs.~(\ref{GME},\ref{Hint}) take the form
\begin{small}
\begin{equation}
\begin{split}
&\frac{d\hat{\rho}}{dt}\!=\!i\left[\hat{\rho},\hat{H}_{int}\right]\!-\!\sum_{j=1,2}\!\gamma_j\!\left(\hat{a}_j^\dag\hat{a}_j\hat{\rho}
\!+\!\hat{\rho}\hat{a}_j^\dag\hat{a}_j\!-\!2\hat{a}_j\hat{\rho}\hat{a}_j^\dag\right)\cr
&\!-\!2\gamma_1N\left[\hat{a}_1^\dag,
\left[\hat{a}_1,\hat{\rho}\right]\right]\!+\!\gamma_1
M\left(\hat{a}_1^2\hat{\rho}\!+\!\hat{\rho}\hat{a}_1^2\!-\!2\hat{a}_1\hat{\rho}\hat{a}_1\!+\!\emph{h.c}\right)\!,\cr
\quad &\hat{H}_{int}=i\sqrt{2}|{\cal
E}|\left(\hat{a}^\dag_1-\hat{a}_1\right)-i\frac{\chi}{4}\left(
\hat{a}_2^\dag\hat{a}_1^2-\emph{h.c}\right). \label{MEsfgS}
\end{split}
\end{equation}
\end{small}
With re-scaling ${\cal E}'\!=\!\sqrt{2}|{\cal E}|,\!\ \!\chi'\!=\!{\chi}/{2}$,
Eqs.~(\ref{MEsfgS}) coincide with the master equations obtained for degenerate SFG, leading to a unified formulation of two-mode SFG for any mode pair, regardless of the frequency separation.

For a fully quantum treatment we apply now the positive
P-representation method \cite{Prep,Gardiner} to Eq.~(\ref{MEsfgS}),
which yields It\^{o} stochastic differential equations (SDE's)
\begin{small}
\begin{equation}
\begin{split}
&\left[\begin{array}{c}\frac{\partial}{\partial
t}\alpha_1\\\frac{\partial}{\partial t}\alpha_1^+\end{array}\right]=
\left[\begin{array}{c}-\gamma_1\alpha_1+\chi'\alpha_1^+\alpha_2+{\cal
E}'\\-\gamma_1\alpha_1^++\chi'\alpha_1\alpha_2^++{\cal E}'
\end{array}\right]+\\
&\ \ \ \ \ \ \ +\left[\begin{array}{cc}\chi'\alpha_2+2\gamma_1M&2\gamma_1N\\2\gamma_1N&\chi'\alpha_2^++2\gamma_1M\end{array}\right]^\frac{1}{2}
\left[\begin{array}{c}w_1(t)\\ w_2(t)\end{array}\right]\\
&\left[\begin{array}{c}\frac{\partial}{\partial t}\alpha_2\\
\frac{\partial}{\partial t}\alpha_2^+\end{array}\right]=
\left[\begin{array}{c}-\gamma_2\alpha_2-\frac{\chi'}{2}\alpha_1^2\\-\gamma_1\alpha_2^+-\frac{\chi'}{2}\alpha_1^{+^2}
\end{array}\right],
\end{split}
\label{PsdeS}
\end{equation}
\end{small}
where $\alpha_1,\alpha_1^+,\alpha_2,\alpha_2^+$ are independent c-number stochastic variables associated with the
field operators, and $ w_1(t), w_2(t)$ are real Gaussian noises, obeying
\begin{equation}
\langle w_k(t)\rangle=0,~~~\langle w_j(t)w_k(t')\rangle=\delta_{jk}\delta(t-t').
\end{equation}

Defining the normally ordered intra-cavity quadratures
\begin{equation}
q_j=\alpha_j+\alpha_j^+,~~~p_j=-i(\alpha_j-\alpha_j^+),\label{Pquadrature}
\end{equation}
and re-scaling with respect to $\gamma_1$ ($\tau\!\equiv\!\gamma_1 t,\gamma_r\!\equiv\!\gamma_2/\gamma_1, \mu\!\equiv\!{\cal E}'/\gamma_1$),
we obtain from SDE's (\ref{PsdeS})
\begin{small}
\begin{equation}
\begin{split}
&\left[\begin{array}{c}\frac{\partial}{\partial
\tau}q_1\\\frac{\partial}{\partial \tau}p_1\end{array}\right]=
\left[\begin{array}{c}2\mu-q_1+{\rm g}\sqrt{\frac{\gamma_r}{2}}(q_1q_2+p_1p_2)\\-p_1+{\rm g}\sqrt{\frac{\gamma_r}{2}}(q_1p_2-q_2p_1)\end{array}\right]+\left[\begin{array}{cc}1&1\\-i&i\end{array}\right]\cdot\\
&\cdot\left[\begin{array}{cc}{\rm g}\sqrt{\frac{\gamma_r}{2}}\left(q_2\!+\!ip_2\right)\!+\!2M&2N\\2N&{\rm g}\sqrt{\frac{\gamma_r}{2}}\left(q_2\!-\!ip_2\right)\!+\!2M\end{array}\right]^\frac{1}{2}\!
\left[\begin{array}{c}w_1(\tau)\\ w_2(\tau)\end{array}\right],\\
&\left[\begin{array}{c}\frac{\partial}{\partial \tau}q_2\\\frac{\partial}{\partial \tau}p_2\end{array}\right]=
\left[\begin{array}{c}-\gamma_rq_2-{\rm g}\sqrt{\frac{\gamma_r}{8}}(q_1^2-p_1^2)\\ -\gamma_rp_2-{\rm g}\sqrt{\frac{\gamma_r}{2}}q_1p_1 \end{array}\right],
\end{split}
\label{qpG}
\end{equation}
\end{small}
where ${\rm g}\equiv{\chi'}/{\sqrt{2\gamma_1\gamma_2}}$
characterizes the strength of the quantum fluctuations internal to
the SFG cavity \cite{LTS}. For any classical amplitude ${\cal E}'$
the validity of Eqs.~(\ref{qpG}) is guaranteed if the ratio of
nonlinearity to damping ${\chi'}/{\gamma_j}$ is small
\cite{Prep,LTS}. Note that Eqs.~(\ref{qpG}) contain two independent
quantum noise contributions: one from the external noise induced by
the coupling to the squeezed light, represented by $M$ and $N$; and
the other - the internal noise proportional to the intra-cavity SFG
field amplitude $\alpha_2=q_2+ip_2$, arising from the non-linear interaction.
An important consequence is that for low SFG efficiency, the
intra-cavity noise may be neglected compared to the squeezed input
noise, which proves crucial for the SFG detection accuracy
calculated below.

Eqs.~(\ref{qpG}) are complicated to be solved exactly. However, if
the input noise is small compared to the
classical terms (a reasonable assumption for an
OPO well above threshold \cite{OPOAT}), linearization methods
can be justified to obtain approximate results
\cite{OptActa28,Strogatz}. Within the zero-order approximation,
corresponding to the classical dynamics, we discard the noise terms
and obtain
\begin{small}
\begin{equation}
\begin{split}
&\frac{d}{d\tau}q^{(0)}_1\!=-q^{(0)}_1\! +2\mu+\!{\rm g}\sqrt{\frac{\gamma_r}{2}}(q^{(0)}_1q^{(0)}_2\!+\!p^{(0)}_1p^{(0)}_2),\\
&\frac{d}{d\tau}p^{(0)}_1\!=\!-p^{(0)}_1\!+\!{\rm g}\sqrt{\frac{\gamma_r}{2}}(q^{(0)}_1p^{(0)}_2\!-\!q^{(0)}_2p^{(0)}_1),\\
&\frac{d}{d\tau}q^{(0)}_2=-\gamma_rq^{(0)}_2-{\rm g}\sqrt{\frac{\gamma_r}{8}}(q^{(0)}_1q^{(0)}_1- p^{(0)}_1p^{(0)}_1),\\
&\frac{d}{d\tau}p^{(0)}_2=-\gamma_rp^{(0)}_2-{\rm g}\sqrt{\frac{\gamma_r}{2}}q^{(0)}_1p^{(0)}_1.
\end{split}\label{qp0}
\end{equation}
\end{small}
The steady-state solution of Eqs.~(\ref{qp0}) (setting all time derivatives to zero) yields
\begin{small}
\begin{equation}
\begin{split}
&p^{(0)}_1=p^{(0)}_2=0,~~~~  q^{(0)}_2=-\frac{{\rm g}}{\sqrt{8\gamma_r}}q^{(0)}_1q^{(0)}_1,\\
&q^{(0)}_1=\frac{-2\sqrt[3]{6}+\sqrt[3]{4\left(9{\rm g}\mu+\sqrt{12+81{\rm g}^2\mu^2}\right)^2}}{\sqrt[3]{9}{\rm g}\sqrt[3]{\left(9{\rm g}\mu+\sqrt{12+81{\rm g}^2\mu^2}\right)}}\xrightarrow[g\rightarrow0]{}2\mu.
\end{split}\label{solqp0}
\end{equation}
\end{small}
\
To investigate the temporal behavior of the system within the first
order approximation we now substitute the zero-order solution
Eqs.~(\ref{solqp0}) into the noisy terms of Eqs.~(\ref{qpG}) and
linearize the equations with respect to the deterministic
part \cite{OptActa28,Strogatz}. This leads to
\begin{small}
\begin{equation}
\begin{split}
&\frac{d}{d\tau}\delta q_1\!=\!-\!\left(1\!+\!\frac{\nu^2}{4}\right)\!\delta q_1\!+\sqrt{\frac{\gamma_r}{2}}\nu\delta q_2\!+\!2\sqrt{M\!+\!N\!-\!\frac{\nu^2}{8}}~w_+,\\
&\frac{d}{d\tau}\delta p_1\!=\!-\!\left(1\!-\!\frac{\nu^2}{4}\right)\!\delta p_1\!+\!\sqrt{\frac{\gamma_r}{2}}\nu\delta p_2\!-\!2i\sqrt{M\!-\!N\!-\!\frac{\nu^2}{8}}~w_-,\\
&\frac{d}{d\tau}\delta q_2\!=\!-\!\sqrt{\frac{\gamma_r}{2}}\nu\delta q_1\!-\!\gamma_r\delta q_2 ,~~~ \frac{d}{d\tau}\delta p_2\!=\!-\!\sqrt{\frac{\gamma_r}{2}}\nu\delta p_1\!-\!\gamma_r\delta p_2,
\end{split}\label{Lqp1}
\end{equation}
\end{small}
where $\nu\!=\!{\rm g}q_1^{(0)}$ is a measure of the SFG efficiency
$\eta_{SFG}\!=\!\nu^2/8$ and
$w_\pm\!=\!\left[w_1\!\pm\! w_2\right]/\sqrt{2}$ represent
white noises.

Eqs.~(\ref{Lqp1}) can be solved in Fourier space to obtain the desired SFG quadratures
\begin{small}
\begin{equation}
\begin{split}
&\delta \tilde{q}_2(\Omega)=-\frac{\nu\sqrt{2\gamma_r}\sqrt{M+N-\frac{\nu^2}{8}}~\tilde{w}_+(\Omega)}{\gamma_r\left(1+\frac{3\nu^2}{4}\right)+i\left(1+\gamma_r+\frac{\nu^2}{4}\right)\Omega-\Omega^2},\\
&\delta \tilde{p}_2(\Omega)=i\frac{\nu\sqrt{2\gamma_r}\sqrt{M-N-\frac{\nu^2}{8}}~\tilde{w}_-(\Omega)}{\gamma_r\left(1+\frac{\nu^2}{4}\right)+i\left(1+\gamma_r-\frac{\nu^2}{4}\right)\Omega-\Omega^2}.
\end{split}\label{qp1}
\end{equation}
\end{small}
Using Eqs.~(\ref{qp1}) and the correlation properties of white noise $\langle\tilde{w}_j(\Omega_1)\tilde{w}_k(\Omega_2)\rangle=\delta_{jk}\delta(\Omega_1+\Omega_2)$ we find
\begin{small}
\begin{equation}
\begin{split}
\langle
&\delta\tilde{q}_2(\Omega_1)\delta\tilde{q}_2(\Omega_2)\rangle=\\
&=\frac{2\gamma_r\nu^2\left(N\!+\!M-\frac{\nu^2}{8}\right)\delta(\Omega_1\!+\!\Omega_2)}
{\gamma_r^2\left(1\!+\!\frac{3\nu^2}{4}\right)^2\!+\!\gamma_r^2\Omega^2\left(1\!-\!\frac{\nu^2}{\gamma_r}\right)\!+\!\Omega^2\left(1\!+\!\frac{\nu^2}{4}\right)^2\!+\!\Omega^4},\\
&\langle\delta\tilde{p}_2(\Omega_1)\delta\tilde{p}_2(\Omega_2)\rangle=\\
&=\frac{2\gamma_r\nu^2\left(N\!-\!M\!+\!\frac{\nu^2}{8}\right)\delta(\Omega_1\!+\!\Omega_2)}
{\gamma_r^2\left(1\!+\!\frac{\nu^2}{4}\right)^2\!+\!\gamma_r^2\Omega^2\left(1\!-\!\frac{\nu^2}{\gamma_r}\right)\!+\!\Omega^2\left(1\!-\!\frac{\nu^2}{4}\right)^2\!+\!\Omega^4},
\end{split}\label{pdc-sfg}
\end{equation}
\end{small}
which express the \emph{intra-cavity} SFG spectral
variances (normally ordered) in terms of the squeezing parameters $N\!+\!M$ and $N\!-\!M$. When $N\!-\!M$ dominates over the internal noise ($N\!-\!M\!\gg\!{\nu^2}/{8}$), the SFG quadratures fluctuations directly provide the degree of input squeezing.

Since measurements are performed on the fields outside the
cavity, we now transform the intra-cavity spectral variances in
Eqs.~(\ref{pdc-sfg}) to extra-cavity spectral variances using the
input-output relation $
\hat{\Phi}_j^{out}\!=\!\sqrt{2\gamma_j^{out}}\hat{a}_j\!-\!\hat{\Phi}_j^{in}$
\cite{GardinerCollett}, where $\hat{\Phi}_j^{out},\hat{\Phi}_j^{in}$
are the outgoing and incoming photon operators external to the
cavity and $\hat{a}_j$ is the intra-cavity operator whose dynamics
we have studied so far. For simplicity, we assume
$\gamma_j^{out}\!=\!\gamma_j$, indicating a lossless cavity
where damping is only due to output coupling. Defining general quadratures $\hat{X}_j^\theta\!=\!\left({\rm
e}^{-i\theta}\hat{\Phi}^{out}_j\!+\!{\rm
e}^{i\theta}\hat{\Phi}^{out^\dag}_j\right)$ the output spectral
variance $V^\theta_j(\Omega)$ can be written as \cite{LTS}
\begin{small}
\begin{equation}
V_j^\theta(\omega)\delta(\omega+\omega')
= 1+\langle \Delta X_j^\theta (\omega)\Delta
X_j^\theta(\omega)\rangle_\textsc{p}, \label{Var}
\end{equation}
\end{small}
where the subscript ${\rm P}$ stands for the normal ordering,
$\Delta X_j^\theta\equiv X_j^\theta-\langle X_j^\theta\rangle$ and
the frequency argument denotes a Fourier transform
$\tilde{X}_j^\theta(\omega)=\int \frac{d t}{\sqrt{2\pi}}{\rm
e}^{i\omega t} X_j^\theta(t)$.

It is essential that the input field $\Phi_1^{in}$, which is associated with the input noise term in Eqs.~(\ref{PsdeS}) for
the intra-cavity fields, is properly included. However, for the SFG
mode 2, which is not explicitly driven by any noise ($\Phi_2^{in}=0$), $V^\theta_2(\omega)$ takes the form
\cite{LTS,GardinerCollett}
\begin{small}
\begin{equation}
V^\theta_2(\omega)\delta(\omega+\omega')=1+2\gamma_2\langle
\Delta\tilde{x}^\theta_2(\omega)\Delta\tilde{x}^\theta_2(\omega')\rangle_\textsc{p},
\label{V2Out}
\end{equation}
\end{small}
where $\tilde{x}^\theta_j$ is the generalized intra-cavity
quadrature. Remembering the re-scaling of time by $\gamma_1$, on substituting
Eqs.~(\ref{pdc-sfg}) into Eq.~(\ref{V2Out}), we obtain the desired relations between the measured extra-cavity SFG
quadratures and the squeezing parameters of the input field:
\begin{small}
\begin{equation}
\begin{split}
&V^{0}_2(\Omega)=1+\\
&+\frac{4\gamma_r^2\nu^2\left(N\!+\!M-\frac{\nu^2}{8}\right)}
{\gamma_r^2\left(1\!+\!\frac{3\nu^2}{4}\right)^2\!+\!\gamma_r^2\Omega^2\left(1\!-\!\frac{\nu^2}{\gamma_r}\right)\!+\!\Omega^2\left(1\!+\!\frac{\nu^2}{4}\right)^2\!+\!\Omega^4},\\
&V^{\frac{\pi}{2}}_2(\Omega)=1+\\
&+\frac{4\gamma_r^2\nu^2\left(N\!-\!M+\frac{\nu^2}{8}\right)}
{\gamma_r^2\left(1\!+\!\frac{\nu^2}{4}\right)^2\!+\!\gamma_r^2\Omega^2\left(1\!-\!\frac{\nu^2}{\gamma_r}\right)\!+\!\Omega^2\left(1\!-\!\frac{\nu^2}{4}\right)^2\!+\!\Omega^4}.
\end{split}\label{Vout}
\end{equation}
\end{small}

\begin{figure}
\centering
\begin{tabular}{cc}
\hspace{-.5cm}\includegraphics[scale=0.37]{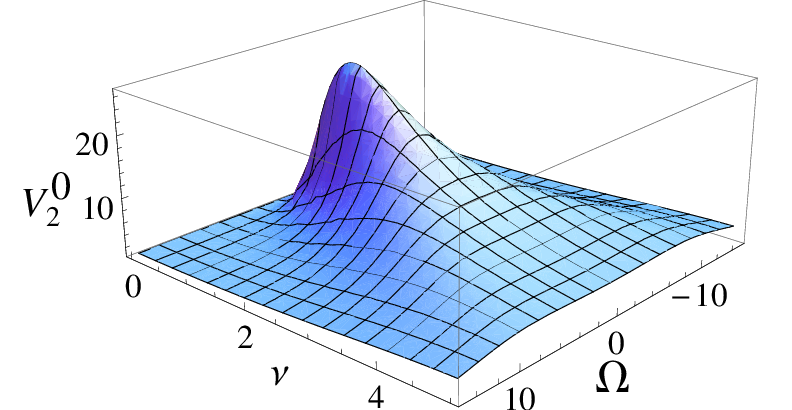}&\hspace{-.9cm}\includegraphics[scale=0.39]{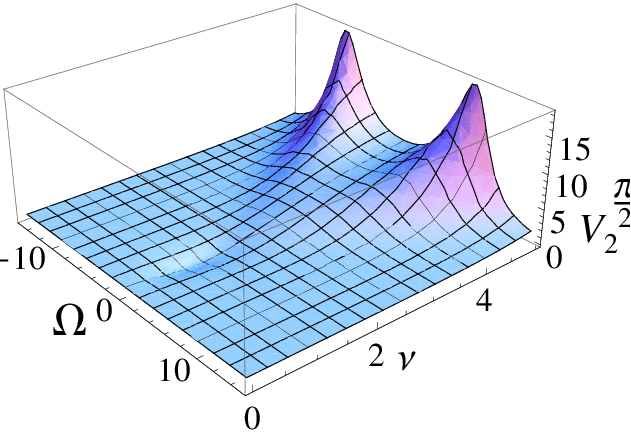}\\
\newline
$(a)\gamma_r=M=10$&$(b)\gamma_r=M=10$\\\\
\newline
\hspace{-.5cm}\includegraphics[scale=0.4]{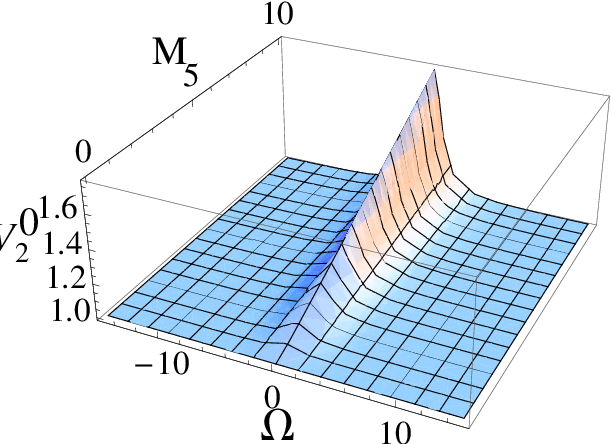}&\hspace{-.9cm}\includegraphics[scale=0.4]{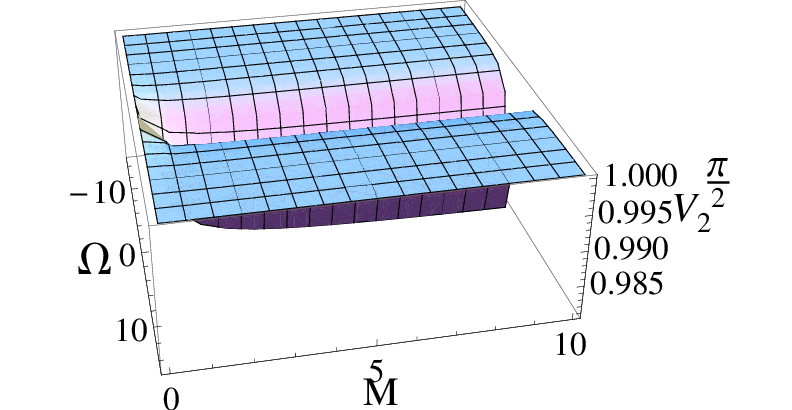}\\
\newline
$(c)\gamma_r=10,\nu=0.1$&$(d)\gamma_r=10,\nu=0.1$\\
\end{tabular}
\caption{\label{SimulationResults} The plot of $V^{0}_2(\Omega)$ Eq.~(\ref{Vout}) [(a) and
(c)] and of $V^{\frac{\pi}{2}}_2(\Omega)$ [(b) and (d)] for ideally
squeezed input $M=\sqrt{N(N+1)}$ as functions of the scaled
frequency $\Omega$, the squeezing parameter $M$, with $\nu={\rm
g}q_1^{(0)}$, ${\rm g}={\chi'}/{\sqrt{2\gamma_1\gamma_2}}$
and $q_1^{(0)}$ given by Eq.~(\ref{solqp0}).}
\end{figure}

By construction within the positive-P representation, Eqs.~(\ref{Vout}) represent a realistic measurement of the extra-cavity SFG quadratures with either partially or ideally squeezed input. The only assumptions are that the SFG cavity is lossless for the input squeezed light (such loss would hinder the squeezing like any other loss) and that the photo-detector efficiency is included in the SFG efficiency $\eta_{SFG}\!=\!\nu^2/8$.
Figure \ref{SimulationResults} illustrates the results of Eqs.~(\ref{Vout}) for an
ideally squeezed input $M=\sqrt{N(N+1)}$ at realistic parameters. As evident, the measured SFG quadratures are either
non-squeezed (in $Q$ quadrature) or undergo insignificant squeezing (in $P$ quadrature). Moreover, for non-ideal squeezed
input, where $N>|M|$, fluctuations of both SFG quadratures are
always above the SQL, which indicates the robustness of the proposed detection to
loss of SFG photons, as sub-SQL fluctuations need not be detected.
We also note that for a non-squeezed input ($M\!=\!N\!=\!0$)
Eqs.~(\ref{Vout}) yield slight squeezing of the SFG output, a known result in SFG cavities
\cite{sSFG}. This squeezing is negligible compared to the external
input noise for low SFG efficiency.

Finally, we can use Eqs. \ref{Vout} to estimate the sensitivity limit of our scheme and the optimal efficiency
$\eta_{SFG}\!=\!\nu^2/8$. Assuming a fast SFG cavity
($\gamma_r^2\!=\!\left(\gamma_2/\gamma_1\right)^2\!\gg\!\Omega^2$),
Eqs.~(\ref{Vout}) can be expanded to 4th order in $\nu$, yielding
\begin{small}
\begin{equation}
\begin{split}
&V^{0}_2(\Omega)\approx
1\!+\!\frac{4\nu^2}{1\!+\!\Omega^2}\left[(N\!+\!M)\!-\!\frac{\nu^2}{8}\left(1\!+\!\frac{3}{4}\frac{N\!+\!M}{1\!+\Omega^2}\right)\right],\\
&V^{\frac{\pi}{2}}_2(\Omega)\approx
1\!+\!\frac{4\nu^2}{1\!+\!\Omega^2}\left[(N\!-\!M)\!+\!\frac{\nu^2}{8}\left(1\!-\!\frac{1}{4}\frac{N\!-\!M}{1\!+\Omega^2}\right)\right].
\end{split}\label{Voutapp}
\end{equation}
\end{small}
Measurement of the squeezing parameters
$N-M$, $N+M$ may be obscured either by vacuum noise (SQL) for low SFG
efficiency (the first $'1'$ term) or by the internal SFG noise for high
efficiency due to input depletion (the last term). The minimum detectable $(N-M)$ is
obtained when all contributions are similar
$4\nu^2(N-M)\!=\!1\ and\ (N-M)\!=\!\nu^2/8$, indicating
that optimal detection occurs for $\nu^2\!=\!\sqrt{2}$ or
$\eta_{SFG}\!\approx\!0.18$, allowing detection down to $(N-M)_{min}\!\approx\!0.18$, very close to the  ideal value $(N-M)_{ideal}\!\geq\!-\!1/2$.
We expect therefore that this new method will find applications in high precision metrology.

This research was partially supported by the Marie Curie IRG program of the European
Union.

\end{document}